\documentclass{article}
\usepackage{graphicx}

\begin{document}

\title{Three particle states and $t$-exchanges}

\author{Norbert~E.~Ligterink\\
Department of Physics and Astronomy, University of Pittsburgh,\\
3941 O'Hara Street, Pittsburgh PA 15260, USA\\
E-mail: nol1@pitt.edu}

\maketitle


\begin{abstract}
The multiple particle final states in
modern data from $4 \pi$ detector experiments offer
a wealth of information about higher hadronic
resonances and hadron interactions. However, it
requires careful analysis to extract model independent
results from those data.
\end{abstract}

\section{Introduction}

The handling of $t$-exchanges in scattering experiments
requires serious consideration. It is the first and
dominant process involving multi-particle states,
which is relevant for the study of hadronic resonances
at energies of the second resonance region ($1.5$ GeV and
above). In most approaches three and more particle states
are handled as effective two-particle states. The state
with one nucleon and two pions is often modeled as a 
rho-nucleon or a sigma-nucleon state. However, if the 
higher resonances are to be understood in a model-independent
way, the three-particle states, and the non-resonant
production processes, such as the $t$-exchange processes
should be handled in their full glory. Fundamentally,
this is not a big problem; the theory is well understood.
However, practically it is a different story. 
Central to the computational problems lie the singularities
associated with the three-particle states going on-shell.
It should be noted that in a simplistic Yukawa treatment 
of the $t$-exchange, ignoring the energy dependence of the
exchange diagram, such a problem does not occur. However,
if the three-particle states and the singularities are 
treated seriously, they should
be regulated and integrated over which requires three scales:
the regulation scale $\epsilon$ $\ll$ the integration
scale $\Delta E$ $\ll$ the physical scale $\mu$. Every problem
is big in numerical sense, because the handling of triple scales. 
Furthermore, if one works with
momentum variables, the singularities form curves in the 
kinematical domain, so extra care is required to locate
the singularities and extrema on the curve with respect to
the chosen grid. In terms
of energy variables the position of the singularity is
obvious; when the scattering energy equals the three-particle
energy, however, the numerical integration still requires great
care. In recent years some methods, such as the Lorentz transform,
have been developed, which essentially takes one away from
the real axis and the singularities, however, they are
rather complicated.

In this paper another approach is investigated. Central is
the expansion of the amplitudes and relevant function in
an orthogonal basis with known dispersion integrals, which
absolves one from doing any singular integration. In terms
of bookkeeping it does require a serious effort, as one
should keep track of the real and imaginary parts of both
the two-particle and the three-particle thresholds and 
amplitudes. The threshold of the three-particle state 
depends on the two-particle states from which it originates;
if the two-particle state has zero momentum, i.e., $E=m_a+m_b$,
the three-particle state it produces can have zero momentum
for all particles too, and its threshold lies at $E=m_a+m_b+\mu$,
while if the two-particle state has a higher momentum, the 
threshold of the three-particle states it can produces will
lie appropriately higher.
Furthermore, given the energies of the initial and final 
two-particle state $\omega_i$ and $\omega_f$, there is an 
upper bound
to the energy of the three-particle state it can produce 
in the $t$-exchange where the exchange particle is emitted
from particle $a$:
\begin{equation}
E_{\rm max/min} = \sqrt{(k_i \pm k_f)^2 + \mu^2} + \sqrt{k_i^2 + m_b^2}
+ \sqrt{k_f^2 + m_a^2} \ \ ,
\end{equation}
where $k$ is the momentum associated with the energy $\omega$:
\begin{equation}
k = \frac{\sqrt{(\omega^2 - m_a^2 - m_b^2)^2 -
 4 m_a^2 m_b^2}}{2 \omega} \ \ .
\end{equation}

Eventually, the kinematical restrictions on a two-particle
state in a given partial wave will lead to a set of 
three-particle states, which can be labeled by the angle between
the momentum of the exchange particle and the momentum
of the two-particle state. This integration can be performed
analytically and leads in the case of a two-particle $s$-wave
to a second-order transition amplitude: 
\begin{equation}
T^{(0)}(E,\omega_i,\omega_f) =
 \frac{g^2}{k_i k_f} \log 
\left[ \frac{E-E_{\rm min}}{E-E_{\rm max}} \right]
\ \ ,
\end{equation}
where $g$ is the coupling constant, and numerical factors
are ignored.
The three-particle state in now implicitly defined through
the imaginary part of the two-particle to two-particle
transition, which saves one from constructing a basis
for the three-particle states. As every three-particle
state follows from the emission of an exchange particle
from the two-particle state, the three-particle
amplitude is given by the two-particle amplitude times
the imaginary part of the elementary transition
amplitude $T^{(0)}$.  The two-particle amplitude is
determined from solving the Lipmann-Schwinger equation
with the elementary transition amplitudes like $T^{(0)}$
in the kernel.

The real part of the amplitude is dominant. The leading order
imaginary part is the first order process given by the 
imaginary part of $T^{(0)}$. However, as with most 
problems, the qualitative changes in the results through
the imaginary parts, i.e., on-shell three-particle states,
can only be understood properly if they are included in
the calculation.

Making a separable expansion of the amplitude lies at the 
basis of an effective method to sum the processes to all order
to yield a unitary and analytic transition amplitude:
\begin{equation}
T^{(0)}(E,\omega_i,\omega_f)  = \sum_{lm} \alpha_{lm}
\phi^{(E)}_l(\omega_i) \phi^{(E)}_m(\omega_f) \ \ ,
\end{equation}
where one should take care the the basis $\phi^{(E)}_a(\omega)$ is
orthogonal (or band-diagonal, such as splines) for a well defined
expansion, and it has known dispersion integrals:
\begin{equation}
\tilde \phi^{(E)}_{lm}(E')= \frac{1}{\pi} 
\int d\omega \frac{\phi_l^{(E)}(\omega)
\phi_m^{(E)}(\omega)}{E' - \omega}
\end{equation}
The real part of the amplitude spans the full kinematical domain
$m_a+m_b <\omega$, hence an expansion in $\xi =
(\omega^2-(m_a+m_b)^2)/\omega^2 \in [0,1]$ can serve as 
a basis $\phi_l(\omega)$:
\begin{equation}
\phi_l(\omega) = \sqrt{4l+2}\frac{(m_a+m_b)}{\omega^{3/2}} 
P_l\left(2\xi-1\right) \ \ ,
\end{equation}
for products of which a closed form dispersion integral can 
be calculated. Therefore, for the expansion of the real part
of $T^{(0)}$ the suffix ``$E$'' is not necessary.

However, the imaginary part of $T^{(0)}$ is only non-zero in a 
restricted 
but infinite domain of $E\otimes\omega_i\otimes\omega_f$; 
when the scattering
energy $E$ lies in between $E_{\rm min}(\omega_i,\omega_f)$ and 
$E_{\rm max}(\omega_i,\omega_f)$.
The domain in the $\omega_i\otimes\omega_f$ space given a
scattering energy $E$ is the energy equivalent of the smooth 
triangular kinematical region of a Dalitz plot, where the each of 
the corners is sharper if the energy is larger compared to 
the mass $m_a$, $m_b$, or $\mu$. As the energies $E$, $\omega_i$,
and $\omega_f$ increase, the domain approaches the triangular
area:
\begin{equation}
E < \omega_i + \omega_f  \ \ \ {\rm and} \ \ \  E>\omega_i
 \ \ \ {\rm and} \ \ \  E>\omega_f \ \ .
\end{equation}
This correlation between the variables does not allow
for a separable expansion of the imaginary part of $T^{(0)}$.

\begin{figure}
\centerline{\includegraphics[width=5cm]{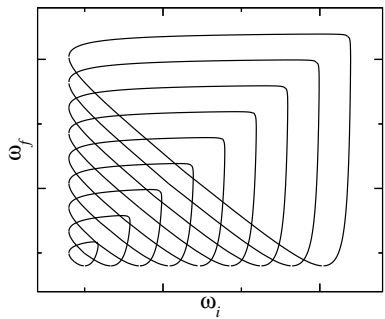}}
\caption{The domains where the imaginary part of $T^{(0)}$ is
nonzero for increasing scattering energies ($m_a=m_b=10\mu$).}
\end{figure}
It seems that a useful parametrization of Dalitz plot amplitudes
and it generalization to more particles and more dimensions
does not yet exist. In the massless case, the domain is a simplex
on which Appell polynomials and generalized mappings of
orthogonal polynomials from the $n$-ball to the $n$-simplex form
a multi-dimensional basis.
In the massive case a system of a weight function that
serves also as domain definition times polynomials $P_I$ in the
single particle energies $\epsilon$ of the three or
more particle state might serve many needs:
\begin{equation}
\phi_I = \theta\left(E - E_{\rm edge} \right) \left(E -
E_{\rm edge}\right)^\beta
P^{(\beta)}_I(\epsilon_1,\cdots ,\epsilon_{n-1}) \ \ ,
\end{equation}
such that 
\begin{equation} 
\int_{{\rm domain}\ E} d\epsilon_1 \cdots d\epsilon_{n-1} 
\phi_I \phi_J = \delta_{IJ} \ \ .
\end{equation}
For each case of different masses ($\pi\pi N,\pi\rho N$, etc)
the coefficients of the orthogonal 
polynomials $P_I$ must be determined. Furthermore, the dispersion
integrals must be calculated for both:
\begin{equation}
\frac{1}{\pi}\int d \omega \frac{\phi_I \phi_J}{E-\omega} \ \ \ 
\ \ \ {\rm and}  \ \ \ \ \ \ 
\frac{1}{\pi}\int d \omega \frac{\phi_I}{E-\omega} \ \ \ .
\end{equation}
However, after this preliminary work, the calculation of
analytic and unitarity transition amplitudes that incorporate
resonant and $t$-exchange contributions, and three-particle
final states will be a straighforward problem. This work
is under investigation.


\begin{thebibliography}{0}
\bibitem{lig03} N.~E.~Ligterink, {\it Phys. Rev. C}
{\bf 67}, 035203 (2003).
\end{thebibliography}
\end{document}